# Influence of Structural Defects on Charge Density Waves in 1T-TaS$_2$

I. Lutsyk[1], K. Szalowski[1], P. Krukowski[1], P. Dabrowski[1], M. Rogala[1], W. Kozlowski[1], M. Le Ster[1], M. Piskorski[1], D.A. Kowalczyk[1], W. Rys[1], R. Dunal[1], A. Nadolska[1], K. Toczek[1], P. Przybysz[1], E. Lacinska[2], J. Binder[2], A. Wysmolek[2], N. Olszowska[3,4], J.J. Kolodziej[3,4], M. Gmitra[5,6], T. Hattori[7], Y. Kuwahara[7], G. Bian[8], T.-C. Chiang[9], and P.J. Kowalczyk[1*]

1 Faculty of Physics and Applied Informatics, University of Lodz, Pomorska 149/153, 90-236 Lodz, Poland

2 Faculty of Physics, University of Warsaw, Pasteura 5, 02-093 Warsaw, Poland

3 Faculty of Physics, Astronomy, and Applied Computer Science, Jagiellonian University, Lojasiewicza 11, 30-348 Krakow, Poland

4 National Synchrotron Radiation Centre SOLARIS, Jagiellonian University, Czerwone Maki 98, 30-392 Kraków, Poland

5 Institute of Physics, Faculty of Science, Pavol Jozef Šafárik University in Košice, Park Angelinum 9, 040 01 Košice, Slovakia

6 Institute of Experimental Physics, Slovak Academy of Sciences, Watsonova 47, 040 01 Košice, Slovakia

7 Department of Precision Engineering, Graduate School of Engineering, Osaka University, 2-1 Yamada-oka, Suita 565-0871, Japan

8 Department of Physics and Astronomy, University of Missouri, Columbia, Missouri 65211, United States

9 Department of Physics and Frederick Seitz Materials Research Laboratory, University of Illinois at Urbana-Champaign, Urbana, Illinois 61801-3080, United States

* corresponding author: pawel.kowalczyk@uni.lodz.pl

## Abstract

The influence of intrinsic defects of 1T-TaS$_2$ on charge density waves (CDW) is studied using scanning tunneling microscopy and spectroscopy (STM, STS), angle-resolved photoelectron spectroscopy (ARPES), and density functional theory (DFT). We identify several types of structural defects and find that most have a local character limited to the single CDW site, with single exception which effectively behaves as a dopant, leading to band bending and affecting multiple neighboring sites. While only one type of defect can be observed by STM topographic imaging, all defects are easily resolved by local density of states (LDOS) mapping with STS. We correlate atomically-resolved STM periodicity of defect-free 1T-TaS$_2$ to top sulfur atoms and introduce tiling of the surface using equiangular hexagon. DFT calculations (with included Coulomb interactions) are used to investigate the electronic structure by introducing sulfur vacancy or substituting sulfur with oxygen. The sulfur vacancy is characterized by metallic properties and is identified as an origin of one of observed experimentally defects. Whereas in the case of the latter, the oxidation of 1T-TaS$_2$ is found to result in the loss of magnetic properties expected in defect-free material.

## Introduction

It is known that defects can alter the properties of materials [1–3]. The prime example is n- and p-doping of silicon which have become the foundation of the modern semiconductor industry [4,5]. With gate sizes downscaled to below 10 nm [6,7] it is expected that new materials have to take over to sustain the global need for electronic device miniaturization [7]. Materials with reduced geometry hold promise in that regard, such as two-dimensional (2D) crystals, which can lead to further miniaturization of electronic components [7–10] and emergence of optoelectronic devices [11,12]. Since in these systems the active layer has a thickness of several angstroms, even a single defect can severely change its properties leading to either improvements or deteriorations. Therefore, it is crucial to understand the naturally existing defects in a variety of 2D materials in order to exploit their properties. When the study of 2D materials is limited by their high reactivity [13], their properties are often investigated through their layered bulk counterparts, in which single layers are held together by weak van der Waals (vdW) forces.

Among the growing family of vdW materials, binary compounds of transition metal dichalcogenides (TMDCs) are one of most interesting ones thanks to the variety of their forms and polytypes. In particular, 1T-TaS$_2$ is of special interest due to its complex phase diagram [14–17] related to periodic layer distortion (PLD) [14] associated with the emergence of charge density waves (CDW), Mott band gap formation and metal to insulator transition at reduced temperatures [17]. It was shown recently that stacking faults leading to different mutual arrangement of PLDs in two-layer 1T-TaS$_2$ are responsible for considerable variations of its electronic structure resolved by STM [18]. Modification of its electronic properties associated with the phase transitions was also achieved by STM pulses [19] or laser irradiation [20–23]. In another study, Raman optical activity was shown in bulk 1T-TaS$_2$ [24] suggesting the presence of chiral phases in this material. Chiral properties were also studied as an effect of Ti doping [25,26]. Moreover, Se doping of TaS$_2$ results in emergence of superconductive properties [27]. Interestingly, sulfur vacancies healed by oxygenation was found to enhance superconductivity in 2H-TaS$_2$ [13].

In this paper, we focus on intrinsic defects in 1T-TaS$_2$ by means of scanning tunneling microscopy (STM) measurements supported by density functional theory (DFT) simulations. Our atomic resolution images of defect free surface allow us to introduce tiling by two triangles and equiangular hexagon, which are strictly related to locations of sulfur atoms in top layer of 1T-TaS$_2$. These results are confirmed by DFT calculations. In turn, our scanning tunneling spectroscopy (STS) measurements shown in form of tunneling conductance (TC, calculated as first derivative of current by bias voltage, i.e., dI/dV) plots, images, waterfall plots allow us to observe a number of CDW shape changes ascribed to electron-electron interactions. In particular, dI/dV images recorded in the vicinity of the Fermi level depict several types of defects with differences in local density of states (LDOS) compared to defect-free 1T-TaS$_2$. One of these defects works as an acceptor, affecting a region of several nanometers in its vicinity. Since the density of this type of defect is relatively large in our samples, the surface potential is highly modulated, leading to considerable fluctuations in the position of the lower and higher Hubbard band (LHB and HHB). Another type of defects associated to sulfur vacancy leads to closure of the Mott band gap and emergence of local metallic properties of 1T-TaS$_2$. The spectroscopic features associated to other types of defects are more subtle and are related to either a slight shift of LHB or CDW amplitude modulation at particular energies in dI/dV data. We attempt to simulate these defects by performing both DFT and DFT+U (which includes Coulomb interactions between Ta 5d orbitals [26,28]) calculations of sulfur vacancies and their substitution by oxygen atoms. Our unfolded band structure calculations are then compared to ARPES, showing good agreement. Based on our calculations, we believe that local metallic behavior observed experimentally is related to sulfur vacancies. We also show that magnetic properties for 1T-TaS$_2$ with oxygen substitutions are considerably hindered in contrast to defect-free material.

# Results

## 1T-TaS$_2$ atomic structure

1T-TaS$_2$ is vdW crystal where single layers are composed of three sheets of atoms: S-Ta-S held by covalent bonds (see Fig. 1a). PLD forms as a result of electron-electron interaction at low temperatures leading to Ta atom shift (see blue arrows in Fig. 1b) [29]. The central, undistorted atom (A-Ta in Fig. 1b) is surrounded by six Ta atoms (labeled B) which are shifted by the same amount toward A-Ta atom.

Six second nearest neighbor Ta atoms (labeled C) undergo even stronger distortions and are also shifted toward A-Ta. These thirteen Ta atoms form a geometric figure called hexagram (from Greek) or sexagram (from Latin), also known as David Star [29] (see Fig. 1b). The geometry of the David Stars allows the formation of CDW/PLD chiral superstructure in two equivalent directions [23,30]. In consequence of PLD formation, the unit cell of 1T-TaS$_2$ (indicated using black rhombus) is considerably larger than in the undistorted case. This results in reduction of the first Brillouin zone as shown in inset in Fig. 1b, requiring unfolding of the calculated electronic band structures (see discussion below).

The PLD is well reproduced in DFT calculations carried out for a single layer of 1T-TaS$_2$. The structural model for single-layer 1T-TaS$_2$ stemming from our DFT calculations is shown in Fig. 1c where top and bottom layer S atoms are indicated using different shades of yellow (pale yellow used for bottom layer). The red arrows show the displacement vector affecting individual S atoms in the unit cell (located in the top layer). Our calculations show that distortions of Ta atoms are mainly in plane in contrast to S atoms which distort in out-of-plane directions. In consequence of this out-of-plane distortion, surface sensitive STM is capable to resolve individual S atoms for which (according to our calculations shown below) the LDOS is higher than on Ta atoms. Note, in the literature, one can find reports showing atomic resolution based on Ta atoms [31], which is most likely related to the STM tip condition [32] resulting in image inversion [33–36]. High resolution STM image of 1T-TaS$_2$ surface is shown in Fig. 1d. It is characterized by clear CDW modulation (as a result of PLD formation), which is further modulated by protrusions corresponding to individual S atoms. Inspection of the CDW centers in Fig. 1d reveals the presence of three protrusions arranged in a triangle (labelled 1). These three protrusions correspond to three top layer S atoms, nearest neighbors of A-Ta, which are shifted out of plane (with highest magnitude indicated by arrows in Fig. 1c; by the plane we understand here original Z plane at which S atoms are located before relaxation). They are indicated in Fig. 1c using yellow rings labelled 1 (in short for 1-S). Note, diameter of the rings shown in Fig. 1c corresponds to relative distance of S atoms measured with respect to Ta plane, i.e. a larger ring diameter indicates S atom located further apart (at the same time this atom is closer to the observer).

Three 1-S atoms in Fig. 1d are surrounded by nine protrusions which form an equiangular hexagon (purple shape in Fig. 1b, c and d) with a longer edge twice the length of the shorter one, i.e. spirolateral $2_{120°}$ [37] characterized by $D_3$ symmetry. In the center of longer edge of this hexagon, a slightly brighter protrusion is measured in STM images (see Fig. 1d). This protrusion corresponds to 2-S atom (see Fig. 1c) which is also pushed above original S plane but with lower magnitude than 1-S. The remaining six S atoms (labeled 3-S) are shifted below the initial S plane and are located in the corners of the equiangular hexagon. Note, the equiangular hexagon is centered with the David Star and its center sits in a corner of the unit cell (see Fig. 1b and c). Inspection of the structural model shown in Fig. 1c reveals that atoms belonging to three neighboring CDWs form an equilateral triangle with three 3-S atoms on each edge (see green triangle in Fig. 1c and d). This triangle contains three C-Ta atoms which are the most distorted. Clearly these atoms are imaged as three dips in STM images (see Fig. 1d). There is no top layer located S atom in the center of this triangle and the protrusion seen in experimental data is a result of overlap of charge related to three 3-S atoms located at its edges. Three remaining C-Ta atoms are imaged by STM as even larger dips and are located in a region indicated using red triangle in Fig. 1. In the center of red triangle, 4-S atom is located. Despite the very large deformation of the nearest neighbor C-Ta atoms, the 4-S atom remains nearly in its original position. It is imaged by STM as a weak protrusion located in the center of red triangle (see Fig. 1d). Both red and green triangles together with purple equiangular hexagon tile the surface of 1T-TaS$_2$. In supplementary information, Fig. S1 shows the equivalent tiling for the bottom surface of single-layer 1T-TaS$_2$ which is different on both sides in this material [24].

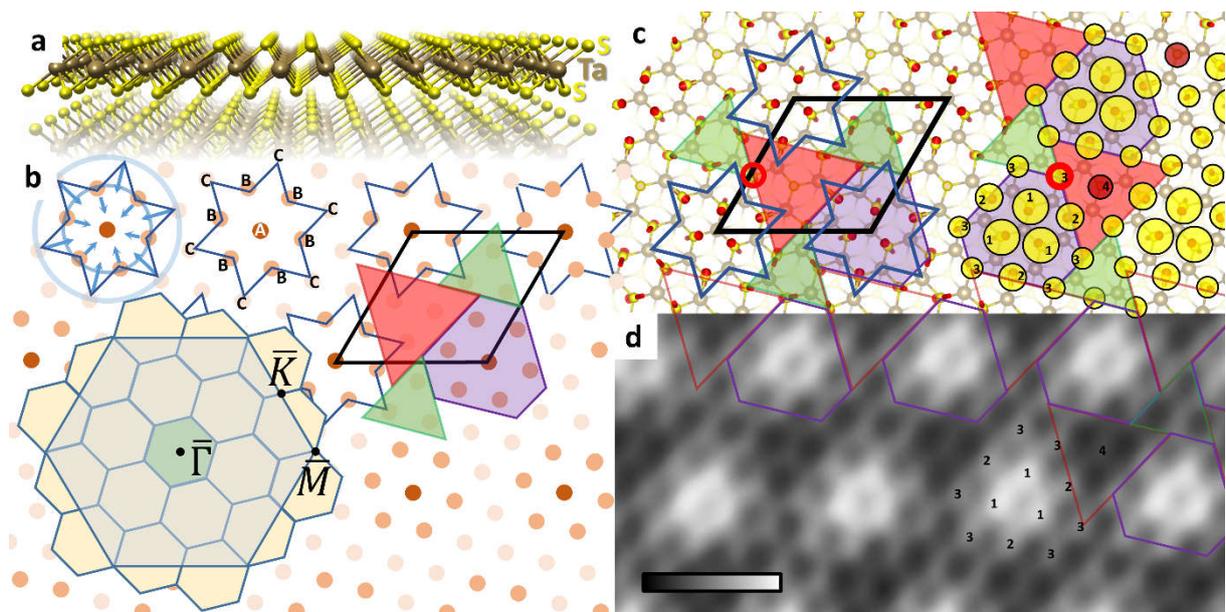

*Fig. 1. (a) Ball and stick model of 1T-TaS$_2$ (side view). (b) Ball model showing deformation of Ta sites as a result of PLD. Ta atoms are indicated using A, B and C labels. Inset shows first Brillouin zone for 1T-TaS$_2$ and for supercell. (c) Ball and stick model of 1T-TaS$_2$ (top view) with top sulfur and middle tantalum atoms shown. Red arrows represent direction and magnitude of top layer S atoms shift in consequence of PLD formation. Yellow and red rings indexed 1-4 indicate locations of S atoms. Radius of rings corresponds to the out-of-plane shift of S atoms. Red ring indicate S site for which DFT calculations are done. (d) High resolution STM image (2 V, 50 pA) showing atomic resolution and CDW recorded at 80 K. Characteristic atomic sites are enumerated 1-4. Scale bar in (d) corresponds to 1 nm. Black (white) corresponds to low (high) topographic height. The red and green triangles together with purple equiangular hexagons represent tiling the surfaces in (b–d).*

## Scanning tunneling spectroscopy studies of defect-free 1T-TaS$_2$

In Fig. 2a shows a STM topography image of the CDW formed on 1T-TaS$_2$. This image was recorded together with a stack of dI/dV images. The CDW shown in Fig. 2a looks uniform on the large area of the sample with only one distinct CDW site (characterized by lower apparent height) indicated D1 in

Fig. 2a. The FFT calculated for this STM image (see Fig. 2b) is characterized by the presence of six features (indicated as q1, q2, and q3), which correspond to real space periodicities. Based on q1-3 features, average distance between CDW can be extracted and for shown here data varies between 1.2–1.4 nm.

Typical dI/dV plot recorded over CDW is shown in Fig. 2c. Its shape suggests $T_A$ type of 1T-TaS$_2$ layer stacking [18] with the Mott gap at the level of 0.4 eV and LHB and HHB located at -0.35 eV and 0.08 eV respectively [29]. Above the HHB, a clear minimum is observed and followed by a plateau extending until the end of scan range, i.e., 0.55 eV. At the beginning of the plateau, a low amplitude maximum labeled B3 is located at approx. 0.29 eV followed by another one located at approx. 0.43 eV. This pattern, i.e., the HHB followed by minimum and plateau with two weak maxima is characteristic for 1T-TaS$_2$ [38].

By examining dI/dV images one can analyze spatial changes in TC data (proportional to LDOS) at constant energy. A few dI/dV images obtained at different energies are shown in Fig. 2d (energy indicated on the bottom right, see also Fig. S2 in supplementary information). It is clear that the CDW periodicity is visible in each of these dI/dV images. Considerable changes in dI/dV mappings are noticeable across the different energies. In particular, for energies up to 0.12 eV the LDOS (proportional to dI/dV) maxima have oval shape and are centered at 1-S atoms (see Fig. 1a). The shape of ovals changes slightly for example at -0.50 eV and -0.33 eV, where initially circular LDOS maxima evolve to elongated ones. A dramatic change is observed between dI/dV recorded at 0.12 and 0.16 eV where within 0.04 eV well-developed protrusions, which are centered at 1-S atoms, are modified into a chicken net-like structure with protrusion moving off 1-S atoms. This structure is further modified as shown in dI/dV image recorded at 0.44 eV (see Fig. 2d). At this energy, dI/dV maxima (minima) are located at centers of green and red triangular tiles (over 1-S atoms) shown in Fig. 1c. In consequence, the hexagonal dI/dV maxima form surrounding central equiangular hexagon with 1-S atoms located exactly in the center. Closer inspection of these data reveals also that slightly larger dI/dV values are located at every second corner of the hexagon (see 0.44 eV in Fig. 2d) and this is attributed to the different atomic arrangement in red and green tiles. Unfortunately, based only on our experimental results, we are not able to determine with certainty if the larger LDOS at 0.44 eV is centered at the red (thus at the location of 4-S atoms) or green tile due to limited lateral resolution during STS measurements. In overall discussed here changes are in agreement to previously published data for similar system of 1T-TaSe2 [39].

Analysis of dI/dV images has the advantage (over dI/dV plots) that subtle differences in TC can be easily observed, even for energy ranges expected to be featureless. Data shown in Fig. 2d and SI Fig. S2 suggests that the LDOS modulation in dI/dV images is also present in the Mott gap and vanishes only for a narrow energy range (-0.02, -0.07) eV (see cyan shaded region in Fig. 2c). The presence of such LDOS modulation is counter intuitive, especially if confronted with dI/dV waterfall plot shown in Fig. 2e recorded along the black arrow labelled p1 in Fig. 2a. Note that weak LDOS modulations are hidden in the Mott gap due to the color scale (see featureless black stripe along cyan marks in Fig. 2e). We solve this issue by normalizing each vertical line of dI/dV waterfall plot (Fig. 2e) to [0, 1] range. Result of this normalization in shown in Fig. 2f and allows to see the dynamics of LDOS periodicities change as a function of energy. Indeed, in the range of (-0.02, -0.07) eV, periodicity is highly distorted (indicating the lack of CDW, which is evident in Fig. 2d at -0.05 eV). The long-range periodicity is also absent just below B3 maximum at 0.24 eV and above B3 at approx. 0.35 eV. Interestingly, a smooth change of LDOS periodicities can be observed for energies above HHB position. The position change continues up in energy toward minimum between HHB and B3. Within a thin narrow energy range, the LDOS periodicities shift their lateral position and change shape from ovals toward chicken net-like pattern. Subtle changes in lateral position of LDOS maxima is also seen below -0.20 eV (see Fig. 2f). These changes are in record with dI/dV images (shown in Fig. 2d and SI Fig. S2). We hypothesize that the observed shifts of the LDOS maxima are related to complex electron-electron interactions and are further analyzed in DFT section below.

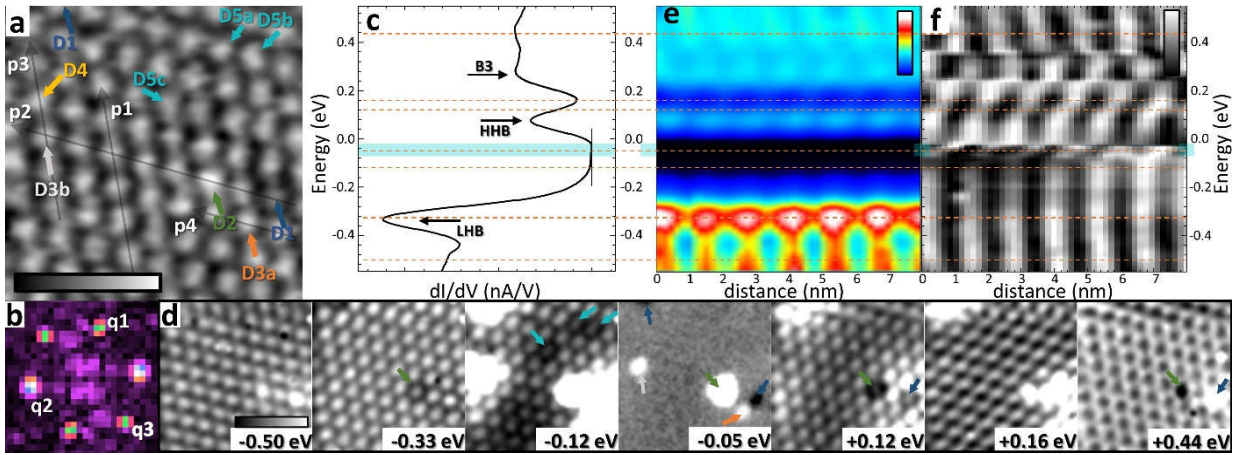

Fig. 2. (a) 11.1×11.1 nm$^2$ STM image (0.55 V, 300 pA) showing CDW on 1T-TaS$_2$ recorded at 80 K. (b) FFT image obtained from (a). Periodicities corresponding to CDW are indicated using labels q1-q3. (c) Typical tunneling conductance dI/dV plot recorded far from defects for $T_A$ stacking of layers in 1T-TaS$_2$. Vertical line indicates zero value of dI/dV. (d) dI/dV images recorded at different energies extracted from dI/dV stack. (e) dI/dV waterfall plot recorded along black line p1 shown in (a). (f) dI/dV waterfall plot after normalization. Small arrows in (a) and (d) indicate location of defects labeled D1-D5. Arrows labeled p1-p4 indicate cross-section lines used to extract dI/dV waterfall plots shown in (e), (f) and Fig. 4 (b)-(d). Horizontal lines in (c), (e) and (f) indicate energies at which dI/dV images shown in (d) were recorded. Shaded region in (c), (e) and (f) shows band gap without clear CDW states. Scale bars in (a) and (d) correspond to 5 nm.

In Fig. 3, we show results of more detailed analysis of the electronic structure landscape, characteristic of 1T-TaS$_2$. Fig. 3a shows both TC dI/dV spectra recorded over the CDW maxima and in the valleys between them. The general shape of both spectra is similar, however TC for CDW maxima has larger LDOS nearly in the whole investigated energy range. This trend changes only for energies larger than 0.36 eV, where TC measured in valleys start to dominate. The TC domination of in-between CDW maxima is also seen in Fig. 2d for 0.44 eV dI/dV image, where CDW is imaged as dark regions and spaces between CDWs has larger TC.

A closer inspection of Fig. 3a reveals also a slight variation of LHB energy location, i.e., LHB measured between CDWs is slightly shifted toward the Fermi energy when compared to LHB position measured on top of CDW. In order to track these changes of the energy location of LHB and HHB, we measured their energy location, which is plotted as spatially resolved maps shown in Fig. 3b and Fig. 3c corresponding to LHB and HHB maximum location respectively (shades of gray represent energy location of the maximum in each pixel of the image). In both maps, one can see that energy position of LHB and HHB change considerably, i.e., by approx. 0.15 and 0.10 eV respectively. We will discuss these changes later. Beside these significant energy shifts, a faint modulation of both LHB and HHB energies is also observed and this modulation clearly is related to CDW seen in the topographic image (see Fig. 2a). The amplitude of the modulation is lower for LHB (see Fig. 3b), where less distinct right diagonal stripes are seen corresponding to spaces between CDWs. This color contrast is in line with dI/dV plot (Fig. 3a) indicating a shift toward the Fermi level of the LHB maxima measured between CDWs. In contrast, the HHB maximum shifts toward the Fermi energy for regions measured at CDW centre and out of the Fermi energy for regions located between CDWs (see pattern shown in Fig. 3c in which spaces between CDWs are lighter, i.e., they are located closer to the Fermi level). These data show that both LHB and HHB maxima shift in registry, i.e., LHB and HHB shift toward and outward the Fermi level respectively due to their location on opposite sides of the Fermi level. This observation is confirmed for Mott gap width ($\Delta_{Mott}$=HHB-LHB) calculated for each pixel and shown in Fig. 3d. At the first sight, the image is relatively uniform without sign of CDW indicating that observed shifts of LHB and HHB are fully correlated. However, a closer inspection allows one to see faint pattern indicating modulation of the Mott gap width. Confirmation of observed modulation of the gap requires more detailed measurements conducted at low temperatures.

In Fig. 3e, we show histograms showing LHB and HHB location. Both histograms are asymmetric with tails located at their right side. These tails are consequences of the LHB and HHB shifts discussed above. In Fig. 3e, we also show a histogram of Mott gap width. The histogram is symmetric, which confirms that shifts of LHB and HHB are correlated. In consequence, the tails observed for both maxima (LHB and HHB) nearly completely cancel and one obtains roughly uniform Mott gap width equal to approx.

0.4 eV observed all over the surface (except for several regions discussed below). The estimated value of $\Delta_{Mott}$ here is in agreement to literature reports, suggesting the existence of a $T_A$ type of layer stacking [18].

### Scanning tunneling spectroscopy studies of defects in 1T-TaS$_2$

As discussed earlier, LHB and HHB amplitude and energy position are changing in record with CDW but there is also long-range modulation seen in Fig. 3b and c. These changes can be better seen in Gaussian blurred images shown in Fig. 3f-h corresponding to LHB, HHB, and $\Delta_{Mott}$ respectively. In these images, extended color scale is used to show aforementioned modulation in a more pronounced way. This allows us to see that both LHB and HHB are characterized by regions with roughly constant energy location of these maxima visualized as uniform shading in Fig.3f and 3g crossing image along its right diagonal (note, color scale for Fig. 3f is inverted in respect to Fig. 3b; bluish and whitish tones in Fig. 3f and Fig. 3g are used to indicate regions with energies close and further away located in respect to the Fermi level respectively). In contrast, the energy locations of both maxima vary in a considerable way along left diagonal (see Fig.3f and 3g). This variation is related to presence of several types of TaS$_2$ defects (labelled D1-5), which locations are indicated by arrows in Fig. 2a,d and Fig. 3b,h. In LHB map (Fig. 3b,f), D1 causes deformation of energy positions of the band (i.e. shift toward the Fermi level) in region of approx. 4 nm around its center. In our opinion, two of such defects separated by 7-8 nm are responsible for the shift toward the Fermi level of LHB in the left side of the map (one of D1s is seen in top left as bluish oval at the top edge of Fig. 3f, the second one is located on the left beyond the scan range). Defect D1 is affecting also HHB causing its shift out of the Fermi level. These two shifts cancel out in regions located few nm away of D1 and in consequence the Mott gap value (Fig. 3d,h) remains constant (compare bottom left corner of Fig. 3f, g and h). In LHB and HHB maps, one can also notice D4 as a spot clearly different from its surrounding (see left side of Fig. 3f and g). D4 seems not to be affected by long-range modification of LHB and HHB (caused by D1 type of defects). Finally, D2 is better resolved in STS maps (see Fig. 3b and c) than in smoothed versions (Fig. 3f and g). However, it is Mott gap map shown in Fig. 3d and 3h in which most of defects can be easily spotted. This indicates that these defects lead to modification of the Mott gap. This modification has rather local nature and is limited to defected CDW site for most observed here defects except D1, which affects a few nanometre-wide region.

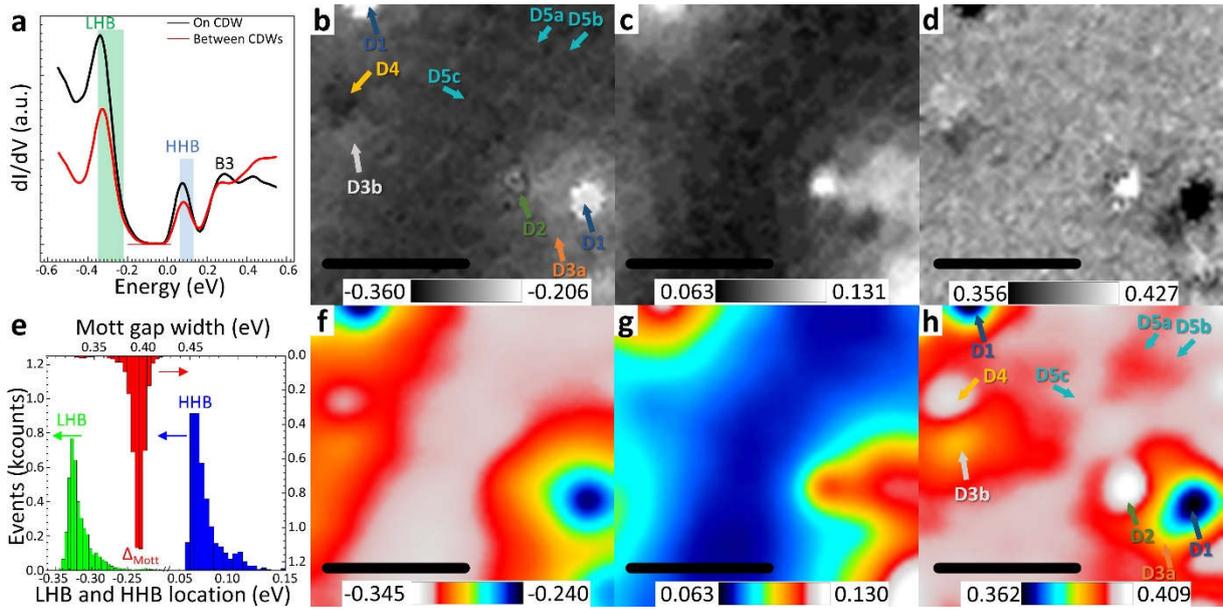

*Fig. 3. (a) Tunneling conductance plot recorded over STM-observed CDW maxima (black) and between them (red). (b) Spatial distribution of LHB energy location extracted from STS measurements conducted over region shown in Fig. 2a. (c) Spatial distribution of HHB energy location. (d) Mott gap calculated as a difference between (c) and (b). (e) Histograms of LHB and HHB locations and Mott gap width. (f), (g) and (h) Gaussian blurred LHB, HHB and Mott gap in (b), (c) and (d) respectively. Green and blue bars in (a) show spread of LHB and HHB respectively. Arrows and labels in (b) and (h) indicate defect locations. Scale bars in (b-d) and (f-h) correspond to 5 nm.*

Further insight in the electronic structure of defects can be extracted from dI/dV plots shown in Fig. 4a and dI/dV waterfall plots extracted along the line profiles of Fig. 2a. In particular, the line p2 crosses three different defects D1, D2, and D3b (see dI/dV waterfall plot in Fig. 4b). At the location of D1, the LHB shifts toward and HHB outward the Fermi level. At the same time, the Mott gap is reduced by 0.04 eV to reach approx. $\Delta^{D1}_{Mott}$ = 0.36 eV. A closer inspection of STM topographic image shown in Fig. 2a reveals that apparent height of the D1-CDW (i.e., CDW maximum associated to D1 defect) is considerably lower than the surrounding ones. In fact, this is the only feature (i.e., apparent height), which allows to spot the presence of D1 defect using STM topographic imaging. The TC at the location the D1 is characterized with a prominent LHB feature and B1 features (B1 is TC maximum located at 0.50–0.60 eV below the Fermi level; for the naming convention see paper by Lutsyk *et al.* [29]), as shown in Fig. 4a, and in the dI/dV waterfall plot shown in Fig. 4b. As discussed above, the presence of D1 affects the electronic structure of the CDWs in its neighborhood, i.e., shift of LHB toward the Fermi level can be observed even 2–3 CDW sites away. Also, HHB is affected and shifted outward the Fermi level but in less extend then LHB. In consequence, the Mott gap is narrower not only at D1 but also in its proximity. This is clearly seen in spatial map of the Mott gap shown in Fig. 3d and 3h. Note, however, that the modification of the gap undergoes within the region smaller then observed shift of LHB and HHB (compare data shown in Fig. 3f-h). This observation is consistent with D1 acting as an acceptor and results in band bending in its neighborhood [40]. Based on the results shown in Fig. 4b, we estimate the surface potential associated with D1 to 0.1 eV, which also indicates the formation of a dipole on the surface [40]. The depletion region extends 4–5 nm away of D1.

The defect labeled D2 has different nature to D1. Most importantly, it has a metallic character with a LDOS peak located 0.03 eV below the Fermi level (see Fig. 4a and 4b). We speculate that this peak is the HHB, which was pushed downward by 0.10 eV from its original position. It is supported by the shape of D2 STS plot (see Fig. 4a) for energies above the Fermi level. One can see three characteristic features typical for this energy range, i.e., minimum and plateau with two maxima. These features are typically preceded by the HHB maximum. These are visible in the D2 STS plot (as mentioned above), and thus attribute these to the HHB. Noticeably, for dI/dV plot recorded over D2 one can also see maximum at roughly the same energy position as LHB in defect-free 1T-TaS$_2$ (shifted up toward the Fermi level by a mere 0.03 eV) but with considerably lower amplitude (see Fig. 4a and b). The observed energy shift is the result of band bending caused by the nearby located D1, consistent with the data shown in the dI/dV waterfall plot (Fig. 4b), where all features in the vicinity of D1-LHB maxima (including D2-LHB) shift toward the Fermi level with a magnitude dependent on the distance to D1. The gap between HHB and LHB is equal to 0.27 eV and is characterized by presence of additional maximum seen as a shoulder approx. 0.12 eV above LHB and 0.16 eV below HHB (maximum location is equal to -0.14 eV). Finally, D2 maximum is characterized by higher STM apparent height than other CDWs (for bias set to 0.55 V). In our experiments, defect is the only one that results in a protrusion in the STM topography data, meaning that tracking such defects in topographic mode is possible by searching for large protrusions.

Cross section line p2 (Fig. 2a) also intersects a defect labeled D3b (see Fig. 4a and b). The same defect is also intersected by the p3 line (see Fig. 4c). Another defect labeled D3a (of the same type) is intersected by the p4 line (see Fig. 4a and d). The general LDOS structure associated with the D3-type defects is similar to that of defect-free 1T-TaS$_2$ (see Fig. 4a and SI Fig. S3). The main difference is the presence of a shoulder at approx. -0.12 eV (see SI Fig. S3 and also bluish region in the Mott gap region in Fig. 4b-d). Since the energy signature of this defect is located in the Mott gap, dI/dV images can be used to reveal and identify them. One of such images recorded at -0.05 eV is shown in the set depicted in Fig. 2d. Both D3 defects are observable as bright (large TC value) regions in the image, i.e., it has nonzero LDOS in the Mott gap and can be treated as an in-gap state. Our experimental results also indicate that the LHB energy at the location of the D3 defect is slightly shifted upward, i.e. toward the Fermi level (see Fig. 3b, 4b, 4c and SI Fig. S3).

The LHB mapping (Fig. 3b) evidences an additional defect type, labelled D4 (see STS plot in Fig. 4a and dI/dV waterfall plot in Fig. 4c). For this defect, a shift of LHB outward the Fermi level is observed, as evidenced by a down-shift of the LHB maximum in the dI/dV waterfall plot (Fig. 4c). Also, the HHB moves toward the Fermi level but with a lesser extent than for the LHB and, in consequence, the Mott gap slightly increases to approx. 0.41 eV. The energy shifts of LHB and HHB, as well as the increase of the Mott band gap can be clearly seen in energy location maps shown in Fig. 3d and 3h. This allows us to hypothesize that in contrast to D1, behaving as an acceptor strongly affecting its neighborhood, D4 has donor properties and has a very limited range of interaction (i.e., in our experiments D4 does not appear to substantially impact its direct environment).

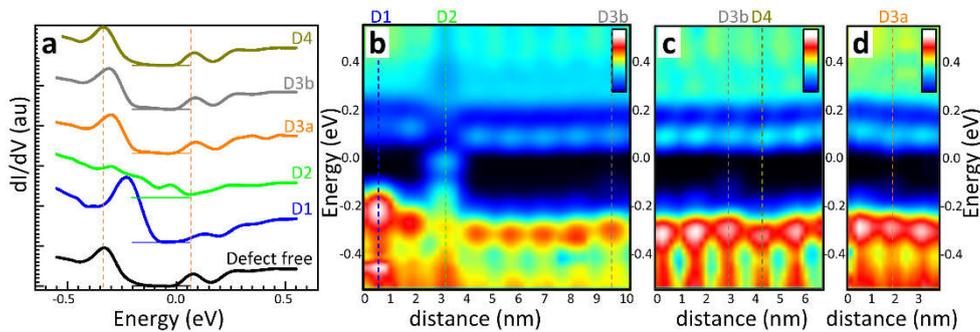

Fig. 4. (a) Tunneling conductance plots recorded over defect-free and defected CDW sites. Plots are shifted vertically. The zero level is indicated by horizontal lines. Tunneling conductance dI/dV waterfall plots extracted along cross-section lines p2, p3 and p4 (shown in Fig. 2a) in (b), (c) and (d) respectively.

Finally, the dI/dV images shown in Fig. 2d also reveal an additional defect, which we label D5. This defect is characterized by a decrease of LDOS in the range -0.09–-0.18 eV (see dI/dV image recorded at -0.12 eV in Fig. 2d). In this energy range, one can see that three of CDW sites (indicated using arrows) are surrounded by darker ovals indicative of a LDOS decrease. Unfortunately, this defect is difficult to characterize in the plots extracted form D5 CDWs and in dI/dV waterfall plots and its exact nature is currently unknown. We speculate that either S3 or S4 atoms are affected, leading to a decrease of the LDOS along a perimeter of the CDW maximum site. Further experiments are required to understand nature of this observation.

## DFT analysis

### DFT of defect-free 1T-TaS$_2$

We now turn to use DFT calculation to gain understanding of the experimental STS data. The reliability of the DFT method is supported by direct comparison of our experimentally recorded ARPES data along the $\bar{\Gamma} - \bar{M}$ direction shown in Fig. 5a and 5b with the DFT-calculated band diagram of single-layer 1T-TaS$_2$ (Fig. 5c). Note, the calculations were carried out using DFT+U formalism [41] (see Methods section, spin up and down bands are indicated using red and blue color respectively) for the 1T-TaS$_2$ supercell, which allows to include the PLD responsible for CDW formation (see Fig. 1b). In consequence of enlarging the unit cell, the 1$^{st}$ Brillouin zone becomes very small (see inset in Fig. 1b) and therefore the calculated bands have to be unfolded to the 1×1 Brillouin zone (see Fig. 5c) before comparison to ARPES data.

Fig. 5a, 5b and 5c showing the experimental ARPES data and unfolded DFT are in very good agreement. Even bands located as far as 8 eV from the Fermi level are reproduced with high fidelity, despite a slight shift toward the Fermi level in the calculation. We attribute this shift to the calculations done for a single-layer 1T-TaS$_2$, while the experiments were carried out on the surface of a bulk 1T-TaS$_2$ crystal. A few band gaps along $\bar{\Gamma} - \bar{M}$ direction in the energy range below 1.3 eV below the Fermi level are also well reproduced both in experimental and calculated data. These gaps are related to PLD formation and strong Coulomb interactions between carriers in 1T-TaS$_2$. Calculation with the inclusion of a Hubbard term leads to spin splitting, which reaches 0.1 eV for bands at $\bar{M}$ approx. 1.3 eV below the Fermi level and at $\bar{\Gamma}$ close to 7 eV below the Fermi level. Larger spin split is observed above the Fermi level reaching 0.15 eV (see Fig. 5c).

The effect of including the Hubbard term on the DOS in the vicinity of the Fermi level is shown in Fig. 5d, in which the calculated DOS for U=0 eV and U=2.27 eV can be compared (see black plots denoted #1 and #1', respectively). Both plots are nearly identical, except for the in-gap state shifting below the Fermi level for U=2.27 eV (indicated by a grey arrow in Fig. 5d). It is clear that the DOS calculated for spin up and spin down are considerably different (see Fig. 5e), which is an indication of magnetism in this system, i.e., calculated total magnetization for 1T-TaS$_2$ is 0.99 $\mu_B$ per supercell. Inspection of the calculated DOS in Fig. 5e shows a clear band gap of approx. 0.4 eV for spin down. For spin up, a band gap also exists but with clear state in the middle, which is a result of spin-splitting of the flat band located at the Fermi level at the $\bar{\Gamma}$ point when Coulomb interactions are not taken into account (see Fig. 5c). The spin down band moves up in energy (see first maximum above the Fermi level for spin down in Fig. 5e) and together with several other bands initially located above the Fermi level forming the HHB state. Note, in Fig. 5e, the Fermi energy is set at the bottom of the conduction band while in Fig. 5d DFT+U calculated DOS is shifted by 0.10 eV (p-doped) in order to align the DOS maxima for the U=0 eV case. In both plots, the LHB, HHB, and B1-B3 states are clearly identified (see shaded areas and black arrows in Fig. 5d) [29]. These spectral features are in reasonable agreement with experimental data (Fig. 3a) and previously published reports [18,29].

Our theoretical data shows that isolated spin-polarized state should form in 1T-TaS$_2$ just below the Fermi level. In the experimental data shown in Fig. 3a, there is no clear signature of such in-gap state. However, we note that the LHB feature fades smoothly towards the band gap as opposed to HHB, which is characterized by a sharper decrease (see Fig. 3a). In our opinion, it is possible that this smooth fading of the LHB into the band gap is a signature of the spin-up in-gap state predicted by DFT+U calculations. In consequence, we expect that this part of the spectrum is strongly spin-polarized which may be detected in spin-polarized STM or spin-polarized ARPES measurements.

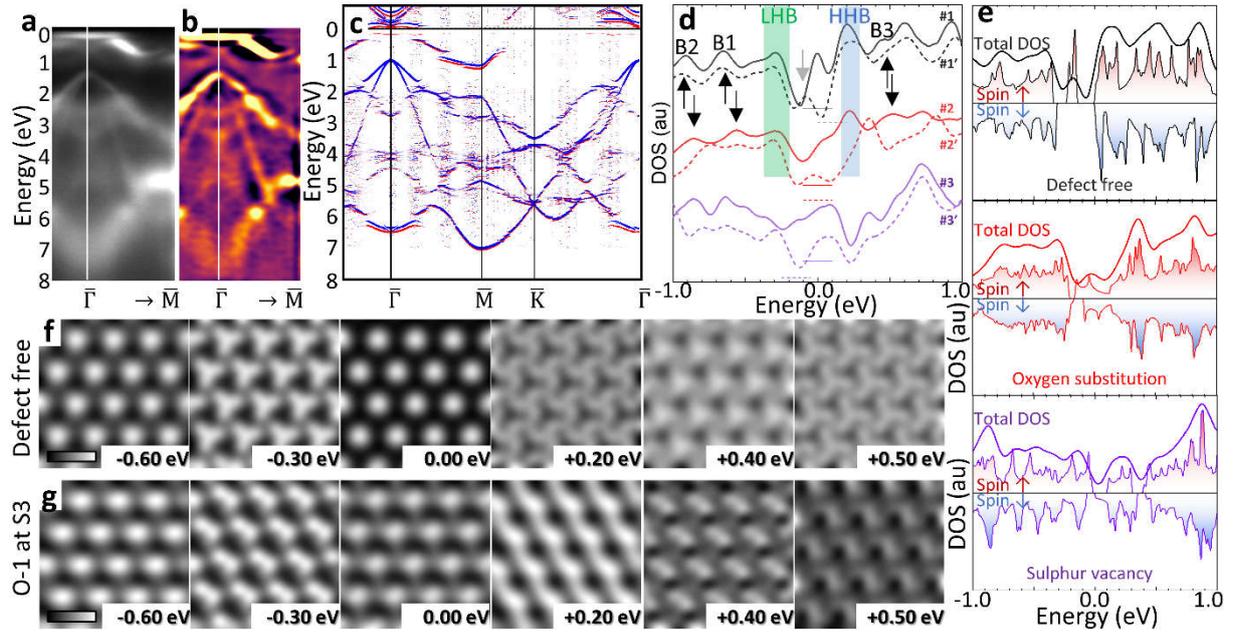

*Fig. 5. (a) ARPES data measured at 80 K along a $\bar{\Gamma} - \bar{M}$ line in k space and (b) its second derivative. (c) Unfolded DFT band diagram for 1T-TaS$_2$ supercell with PLD. Spin up and spin down indicated using red and blue respectively. (d) DFT calculated DOS for defect-free (#1), oxygen substitution (#2) and sulfur vacancy (#3) (see Fig.1a for location of sulfur atom which was substituted and removed). Dashed lines are used to plot DOS calculated using DFT+U. (e) DFT+U calculated spin-polarized DOS and total DOS for (from top) defect free, oxygen substitution and sulfur vacancy. (f) and (g) DFT calculated LDOS images for defect-free and oxygen substitution respectively. Scale bars in (f) and (g) correspond to 1 nm.*

Our experimental dI/dV images of the CDW discussed above indicate a strong dependence on the energy (see Fig. 2d). Our DFT-calculated LDOS shown in Fig. 5f confirm this observation. In order to qualitatively compare these data with STS-recorded dI/dV images, a gaussian filtering operation has been performed on the calculated LDOS images to mimic the effect of tip broadening (LDOS images for an energy range of ±1 eV with 0.10 eV interval are shown in Fig. S4, and corresponding smeared versions are shown in Fig. S5). Below the Fermi level, the appearance of the CDW evolves from oval-shaped at -0.60 eV to a more triangular arrangement at -0.30 eV. In turn, above the Fermi level, a contrast inversion occurs in particular at the HHB location (compare -0.3 eV and +0.20 eV image in Fig. 5f). Both spatial arrangement change and contrast inversion are observed in the experimental data (see the description above). Finally, in the experimental STS image obtained at 0.44 eV above the Fermi level, hexagonal rings can be observed with particularly pronounced corners (see Fig. 2d). This feature is also reproduced in DFT data for 0.40 and 0.50 eV above the Fermi level where brighter sites and characteristic mesh forms, respectively (see Fig. 5f).

Defected 1T-TaS$_2$

In order to understand defect formation in single-layer 1T-TaS$_2$, we decided to simulate sulfur substitution by oxygen and sulfur vacancy as potentially probable point defects (see SI Fig. S6 for structural models). The DFT calculated DOS for such defects are shown in Fig. 5d and are denoted #2 and #3 respectively. The sulfur atom which we substitute (remove) is indicated by red ring in Fig. 1a and denoted S3/1 (see also SI Fig. S6a for locations of two other S sites for which calculations were performed). It is clear that oxygen substitution does not alter the DOS in substantial way. All spectral features, i.e., HHB, LHB, B1-3 are easily resolved and their location in energy does not substantially change with respect to defect-free 1T-TaS$_2$. This behavior is in agreement with previous oxidation studies of the 2H-TaS$_2$ polytype [13]. It is, however, noticeable that for oxidezed 1T-TaS$_2$ the DOS has finite value at the Fermi level, related to the closing of the band gap. Essentially the same result is obtained if DFT+U is used (see middle panel in Fig. 5e and plot #2' in Fig.5d), i.e., all main electronic features are preserved with the exception of the closing of the band gap . Moreover, the +U correction allows us to see that the introduction of an oxygen substitution results in a considerable reduction of the spin polarization of 1T-TaS$_2$, as shown in Fig. 5e, where little differences can be observed in the spin-specific DOS (the calculated total magnetization is 0.17 $\mu_B$ per supercell, which is considerably lower than in defect-free 1T-TaS$_2$, i.e. an almost 6-fold reduction).

The observed changes of the DOS for oxygen substitution case are also clearly visible in LDOS images shown in Fig. 5g (and SI Fig. S4 and S5). CDW shape change is a direct consequence of PLD changes, which are shown in Fig. S6. In consequence, one can search for defect by looking at variation of CDW contrast or its shape change when STM imaging is performed, in particular in LDOS mapping mode.

In contrast to oxygen substitution, the DOS for the sulfur vacancy system is completely different to defect-free 1T-TaS$_2$ (see #3 in Fig. 5d). It is characterized by a maximum located at the Fermi level, which makes it metallic. Moreover, it is characterized by the presence of DOS minima at the location of LHB and HHB, where first one is rather subtle, while the latter is very pronounced. DFT+U calculated DOS is denoted #3' and is shown in Fig. 5d. It is qualitatively very similar to DOS calculated without +U term. It is worth to mention that the spin polarization for sulfur vacancy is relatively strong and comparable to defect free 1T-TaS$_2$ (see bottom panel in Fig. 5e, total magnetization is 1.00 $\mu_B$ per supercell).

The characteristic spectral features for oxygen substitution as well as sulfur vacancy show up also for other sites in the supercell. Two other sites have been examined and the results of our calculations are shown in SI Fig. S6 in which defect-free DOS is compared to two groups of DOS plots, i.e., oxygen substitution and sulfur vacancy. The DOS plots for each type of point defect show that they are qualitatively characterized by the same set of spectral features, with little quantitative difference. This strongly indicates that the two types of defects simulated here have a universal character, independent of the specific sulfur site within the supercell.

LDOS in function of distance

To elucidate whether the STM measurements "see" the sulfur or the tantalum atoms, we calculate the LDOS at different distances from the top sulfur plane of 1T-TaS$_2$. The DFT-calculated LDOS maps are shown in Fig. 6a, for a distance of -0.1, 0.0, 0.1, 0.2 and 0.8 nm above the top S plane. A negative distance indicates that the LDOS is calculated below top S plane, i.e., inside the single-layer of 1T-TaS$_2$. Note, calculations shown in Fig. 6 were conducted for energy window of 0.1 eV in width, centered at 0.0 eV (i.e., at Fermi level). These data clearly show that LDOS measured above the surface is dominated by sulfur atoms (see equiangular hexagon at 0.1 nm plane in Fig. 6a). The contribution of the Ta atoms is dominating below the surface, i.e., inside the layer 1T-TaS$_2$ (see David Star at -0.1 nm plane in Fig. 6a) and is vanishing above the surface (see arrow in Fig. 6a indicating weak LDOS associated to A-Ta atom).

The same trend is preserved also for LDOS calculated for different energies. In Fig. 6b, we show LDOS plotted for A-Ta and 1-S atoms in function of distance and energy. For all energies high LDOS for A-Ta atom is observed below surface. In turn, above the surface of 1T-TaS$_2$, sulfur atoms start to dominate.

In order to compare the LDOS measured over A-Ta and 1-S, we calculate the ratio of both values. Results are shown in Fig. 6c as logarithm of 1-S to A-Ta ratio. This plot shows in a clear way that close to the surface LDOS calculated over 1-S atoms is over ten times larger in comparison to LDOS over A-Ta atom. Further away from the surface the ratio is quickly decreasing approaching 1. This is a consequence of the LDOS smearing, associated with the gradual loss of details in LDOS images as the distance from the surface grows. The loss of detail is highlighted with the LDOS image calculated for distance set to 0.8 nm (see Fig. 6a). In this image, the CDW with a very low LDOS intensity can be seen without any details resulting from individual atomic contribution.

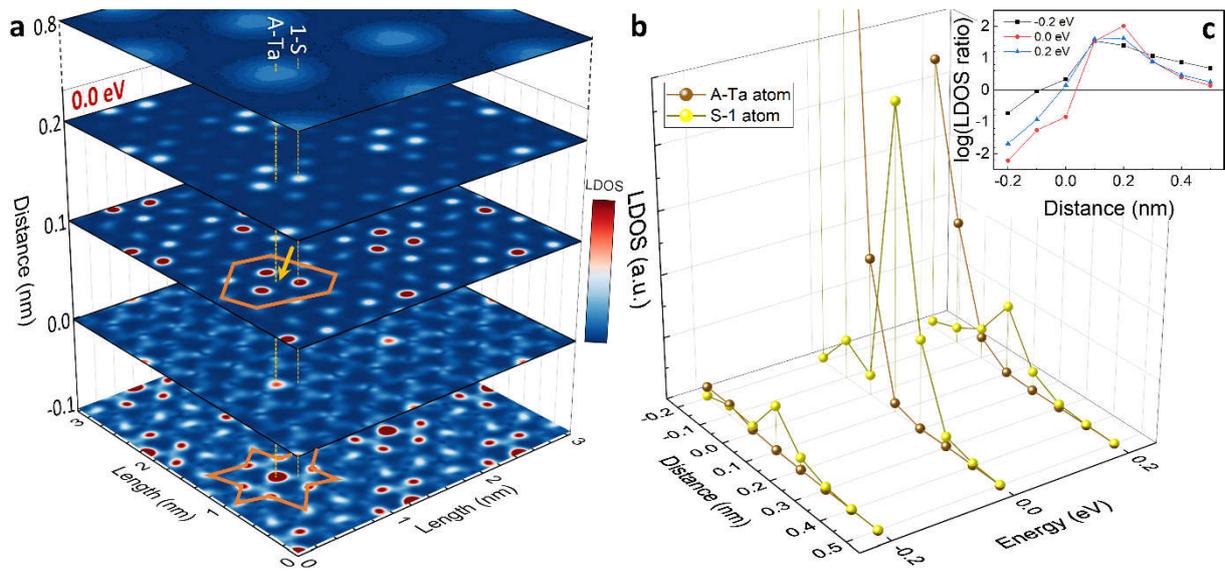

*Fig. 6. (a) Set of LDOS images calculated in function of distance measured from top sulfur atoms plane for energy equal to 0.0 eV. No Hubbard term is used in calculations. David Star and equiangular hexagon are indicated using orange shapes. A-Ta and 1-S atoms are indicated using dashed line. Scale for 0.8 nm LDOS image is a few hundred times exaggerated in comparison to other images in the plot. (b) Plot of LDOS in function of distance and energy for A-Ta and 1-S atoms. (c) Ratio of LDOS measured over 1-S and A-Ta atoms for energies -0.2, 0.0 and 0.2 eV. Negative distance values indicate that the calculations were performed inside 1T-TaS$_2$ (under top layer).*

## Comparison of DFT results with experimental data

The comparison of our DFT results with experimental measurements allows to understand the origin of some of the observed defects. In particular. We find similarities between the DOS plots for sulfur vacancy and D2 type of defect. The experimental band gap (i.e., global minimum) is shifted roughly to the location of HHB of defect-free 1T-TaS$_2$, at the Fermi level pronounced maximum is observed and weaker local minimum coincides with LHB. A similar DOS is observed for sulfur vacancies which suggest that this is the origin of D2 defects.

Our experimental results show that among all defects investigated in this study, the D1-type defect impacts the electronic structure of 1T-TaS$_2$ in a most severe way. We have not been able to reproduce the observed changes of LDOS in our calculations and therefore we speculate that this defect is related to foreign atoms serving as dopants. Previous studies indeed showed that comparable impacts on the LDOS can be observed as described here i.e., shift of HHB and LHB toward the Fermi energy [25,26] which is a strong argument for the presence of adatom in 1T-TaS$_2$. Our XPS measurements (not shown) do not allow to clearly evidence the presence of a foreign atomic species that could be responsible for the origin of such defect.

The influence of D4 and D5 defects on the electronic structure is very subtle and can be observed only in dI/dV waterfall plots and images for certain energies. The origin of these defects is currently unknown and further investigations will be necessary for further identification.

Finally, in our experiments, we achieve atomically-resolved imaging of the 1T-TaS$_2$ surface (see Fig. 1d). We interpret the observed atomic protrusions originating from the sulfur atoms. This claim is supported by our DFT calculations, depicted in Fig. 6, clearly show that above the 1T-TaS$_2$ surface, it is the sulfur atoms that provide the main contribution to the LDOS, for the energy range investigated. The apparent height in the STM images is proportional to the modulation of LDOS at constant energy and the largest fluctuation of the LDOS in our calculations is observed above the sulfur atoms. Thus, we believe that in STM experiments mainly the top-surface sulfur atoms are observed. For larger distances from the surface, the subtle details of the atomic resolution are lost, and the image only resolves the CDW without fain details of atomic structure. Our data also suggests that LDOS associated with Ta atoms can be detected above surface, in particular for the central A-Ta atom. However, LDOS calculated above A-Ta is over ten times weaker than that originating from the 1-S atom. Nonetheless, we cannot exclude the observation of Ta atoms in STM experiments but in such case we expect that the image would be dominated by the protrusions related to sulfur atoms.

## Conclusions

In conclusion, we evaluate the impact of structural defects on the 1T-TaS$_2$ electronic structure, leading to substantial changes of its local and potentially also global properties. Five types of defects (D1–D5) are identified and their electronic structure is characterized using STS. The D1 defect decreases the Mott gap and behaves as an acceptor, leading to band bending in a region up to 5 nm away of its location. Our results show that few such defects located at some distance result in substantial modulation of the lateral electron density landscape, leading to mutual shifts of the lower and higher Hubbard bands. In contrast, the D2 defect is characterized by a metallic STS spectra, whose impact is limited to one CDW site only. Our DFT simulations indicate that it is consistent with a sulfur vacancy. The D3 type of defect is characterized by a slight decrease of the Mott band gap and an emergence of additional electronic states below the Fermi energy. The defect D4 is characterized by a slight increase of the gap. These observations clearly indicate that the band gap width depend on the interplay between layer stacking and on the structural defects of the 1T-TaS$_2$ surface. Finally, the presence of the D5 defect can be spotted in an energy location at approx. 0.12 eV below the Fermi level as a depletion of electronic states around defected CDW sites. All these defects are too subtle to be observed in topographic STM scanning mode. However, the D5 defects become visible when the electronic structure is mapped over an energy range close to the Fermi level. This is supported by DFT calculations, in which the CDW associated with defected supercells is considerably different to the CDW in a defect-free material.

Finally, our DFT calculations show that the predicted magnetism for 1T-TaS$_2$ single layer is substantially quenched if even one sulfur atom is exchanged with an oxygen in supercell. This might be major obstacle that needs to be overcome for experimental magnetism investigations of this material. In contrast, the sulfur vacancies do not affect magnetic properties of 1T-TaS$_2$ in a substantial.

Overall, the defects observed in this paper shed light on the fragility of the 1T-TaS$_2$ electronic structure. The perturbations of this structure are mainly observed within one, affected CDW site. One exception here is the defect working as dopant, disturbing the electronic structure in range of few nanometers. Such naturally existing defects can considerably affect the global properties of 1T-TaS$_2$ and their existence should be taken into account if 1T-TaS$_2$ is considered in future electronic and optoelectronic devices.

## Experimental and theoretical methods

The STM/STS measurements (Unisoku Co., USM1400 controlled by Nanonis BP 4.5) were carried out in UHV at 80 K. Typical pressure during measurements was lower than $1\times10^{-10}$ mbar. The STM tips were prepared by mechanical cutting of 90/10 Pt/Ir wires. The I(V) curves were recorded simultaneously with a constant current image by the use of the interrupted-feedback-loop technique. Based on these measurements, the first derivative of the tunneling current with respect to voltage (dI/dV) was calculated numerically.

ARPES measurements were conducted at Polish National Synchrotron SOLARIS at URANOS (former UARPES) beamline, using a DA30L Scienta-Omicron electron spectrometer. The photon energy was set to 67 eV. Sample temperature during measurements was kept at 80 K.

High-quality 1T-TaS$_2$ crystals were supplied by HQ Graphene and cleaved in situ in UHV condition ($1\times10^{-8}$ mbar) at room temperature.

First-principles calculations were performed using QUANTUM ESPRESSO suite [42,43] implementing the DFT formalism with a plane wave basis [44]. The exchange correlation potential of Perdew–Burke–Ernzerhof (PBE) type [45] and projected augmented wave (PAW) approach [46] were utilized, with a kinetic energy cut-off set to 53 Ry. In the calculations, scalar relativistic (DFT+U) and fully relativistic (DFT) pseudopotentials permitting non-collinear calculations with spin-orbit effects were applied [47]. To capture the coulombic correlation effects for the electrons at 5d orbitals of Ta atoms, spin-polarized calculations within DFT+U formalism were performed on the basis of the rotational-invariant approach after Ref. [48]. The Hubbard U value of 2.27 eV was accepted, as derived from linear response-based calculations [24]. To capture the CDW phase in TaS$_2$, a $\sqrt{13} \times \sqrt{13}$ R13.9º supercell was constructed. In order to model a monolayer system, a slab geometry with at least 15 Å of vacuum separating the periodic images in the direction perpendicular to the layer was used. The 6×6 mesh of k-points was used for relaxation of the atomic positions and self-consistent calculations, whereas a denser mesh of 24×24 (DFT+U) or 48×48 (DFT) was applied to non-self-consistent calculations for DOS determination. In the calculations, the vdW correction was applied [49,50] together with dipole correction introduced in Ref. [51]. The geometry relaxation was performed using a quasi-Newton algorithm. Spatially-resolved LDOS calculations correspond to the plane at constant height of 3 Å over the topmost sulfur atom unless stated otherwise. To permit the comparison of band structures resulting from the supercell-based calculations and from the ARPES measurements, the unfolding procedure from the supercell first Brillouin zone to the normal cell first Brillouin zone was performed with BandUPpy code [52–54].

## Acknowledgments

This work has been supported by the National Science Centre, Poland under grants 2015/19/B/ST3/03142 and 2019/32/T/ST3/00487. M.L.S. thanks for University of Lodz support under IDUB 6/JRR/2021 project. T.C.C acknowledges support of the U.S. Department of Energy, Office of Science, Office of Basic Energy Sciences, Division of Materials Science and Engineering, under Grant No. DE-FG02-07ER46383. M.G. acknowledges financial support provided by Slovak Research and Development Agency provided under Contract No. APVV-SK-CZ-RD-21-0114 and by the Ministry of Education, Science, Research and Sport of the Slovak Republic provided under Grant No. VEGA 1/0105/20 and Slovak Academy of Sciences project IMPULZ IM-2021-42 and project FLAG ERA JTC 2021 2DSOTECH. Author acknowledge the provision of the Polish Ministry of Education and Science project: "Support for research and development with the use of research infrastructure of the National Synchrotron Radiation Centre SOLARIS" under contract nr 1/SOL/2021/2. We acknowledge the SOLARIS Centre for the access to the Beamline URANOS (former UARPES), where the measurements were performed.

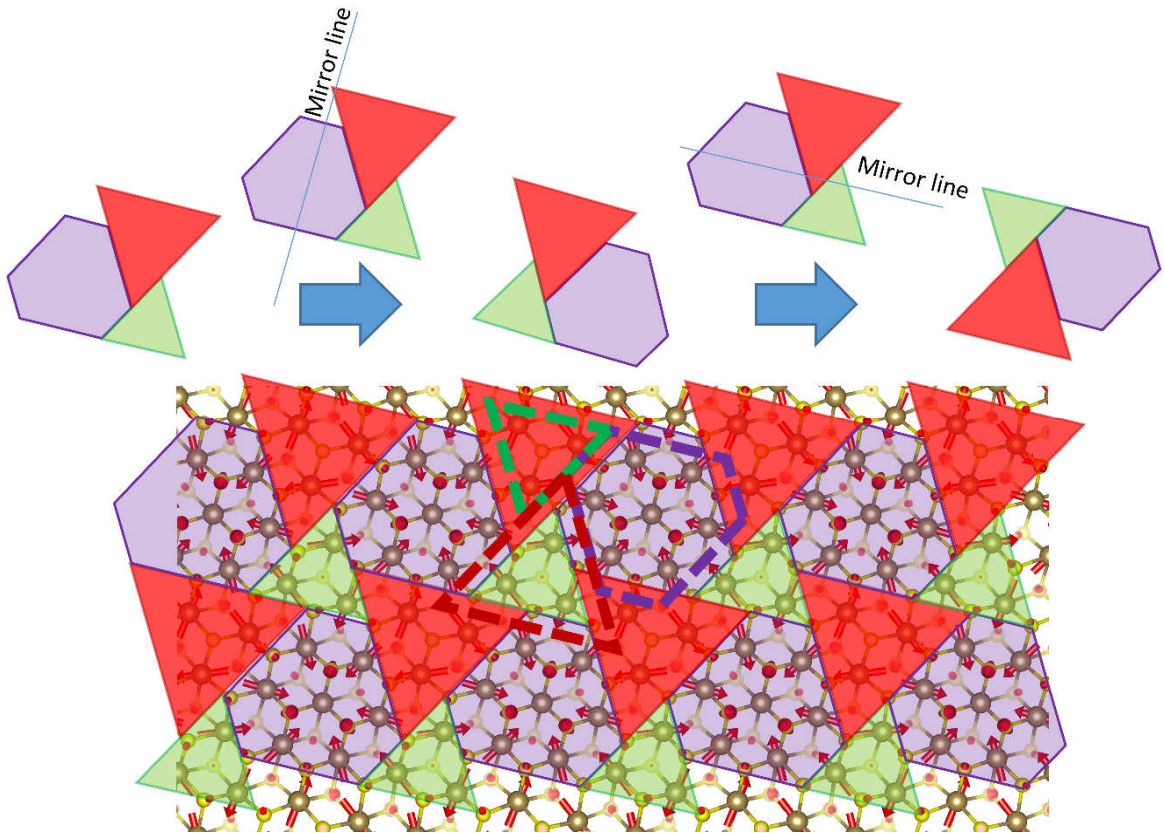

Fig. S 1. Tiling on the bottom surface of 1T-TaS$_2$ requires performing two reflections along planes intersecting purple equiangular hexagon through 1-S, A-Ta, 2-S and perpendicular plane intersecting B-Ta, A-Ta, B-Ta (top panel). Two mirror planes perpendicular to the surface and each other are shown. Mirror reflection of the top tails (left) results in bottom tiles (right). In bottom panel, ball model is shown covered by top tiles. Dashed line is used to show orientation of bottom tiles.

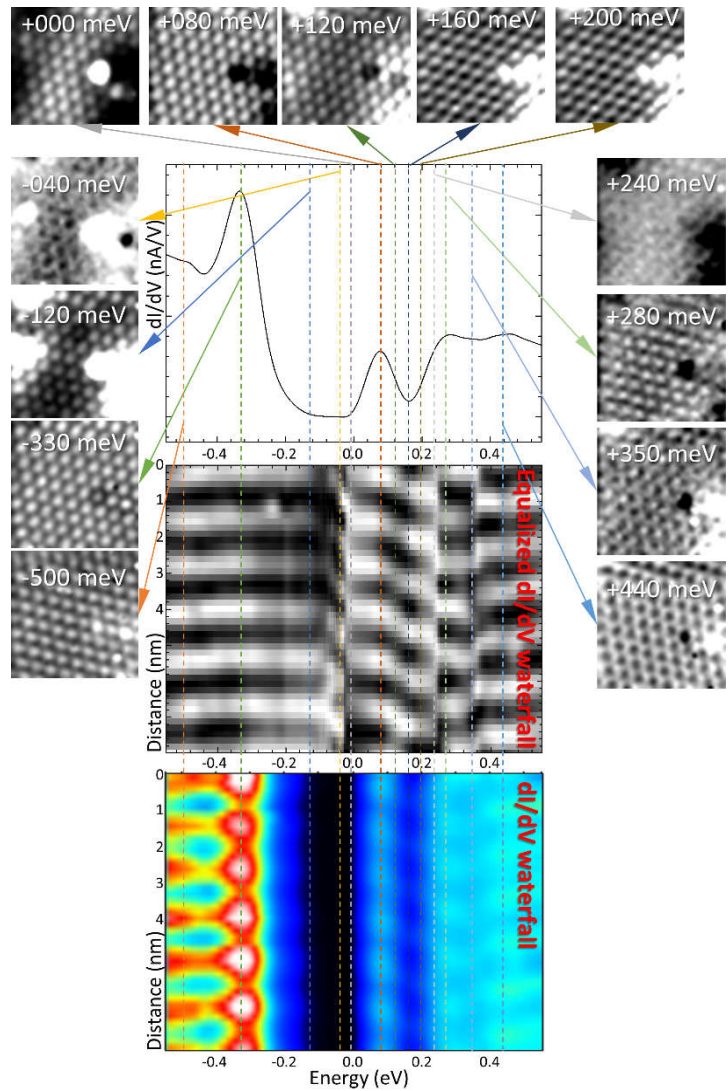

Fig. S 2. dI/dV spectra recorded over CDW on 1T-TaS2 together with tunneling conductance maps equalized and as measured along p1 line shown in Fig. 2 in main part of the paper. Individual tunnelling conductance images recorded at constant energy are shown visualizing changes of the electron density as a function of energy.

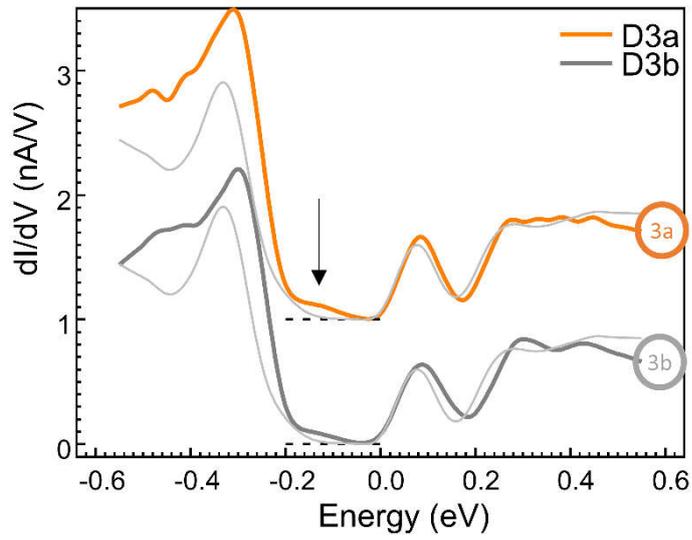

Fig. S 3. Two STS plots recorded over D3a and D3b defects. Arrow indicates presence of additional maximum in the Mott gap, which is characteristic feature for this type of defect. Light grey plots correspond to STS data recorded on defect-free 1T-TaS$_2$. Lift of the LDOS in the region of the Mott gap is clearly seen for both D3a and D3b defects. Black dashed lines correspond to zero.

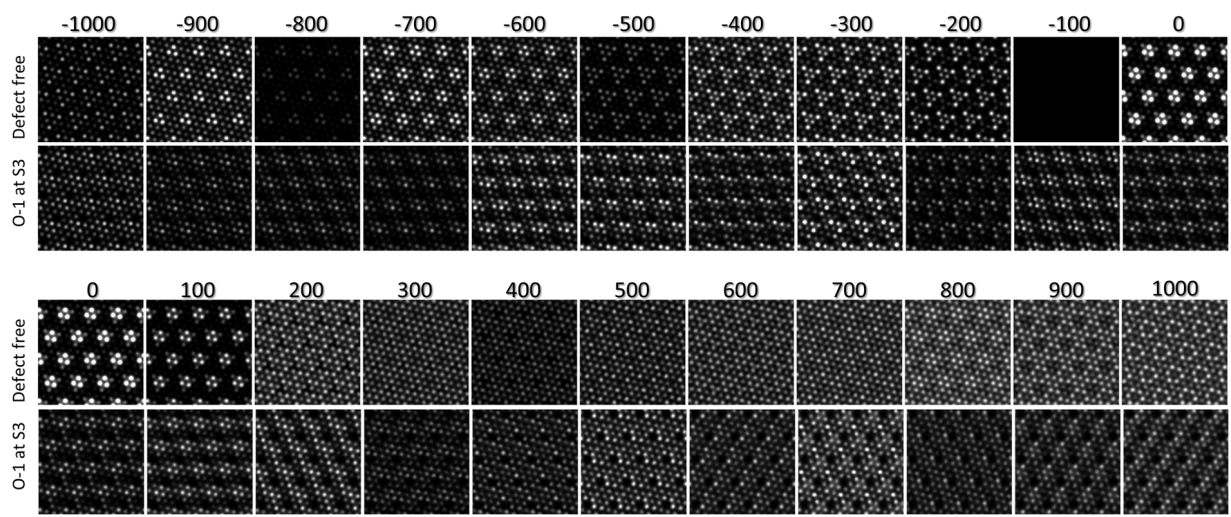

Fig. S 4. DFT calculated LDOS images for energies in range ±1000 meV with 100 meV interval. The energy is given above images in meV. Two rows correspond to defect-free 1T-TaS$_2$ (top row for each energy) and oxygen substitution at S3/1 site (bottom row). These data qualitatively show that CDW formed as a result of PLD formation in pristine and oxidized surfaces are different.

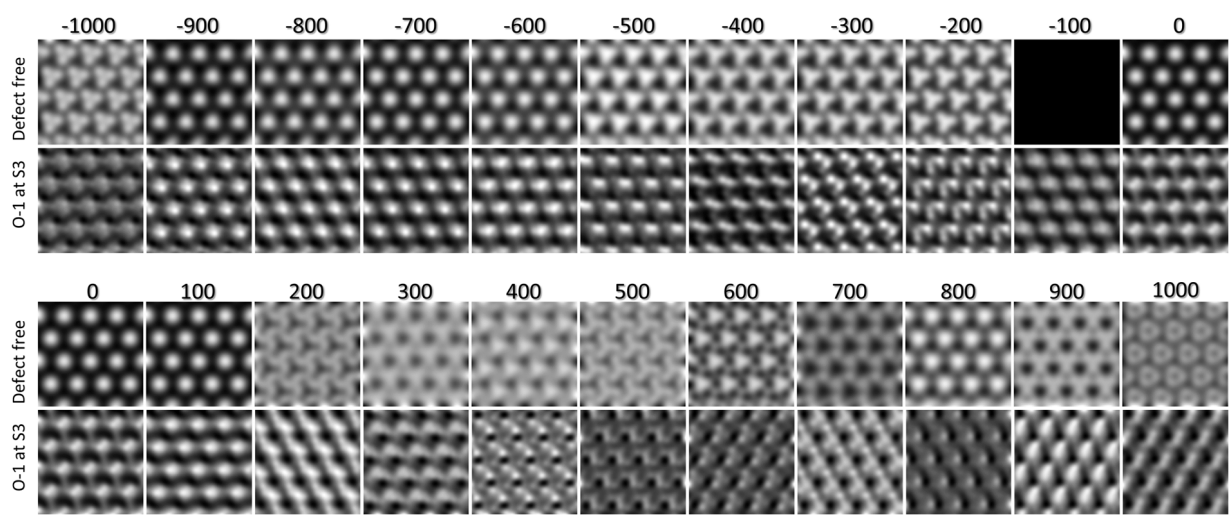

Fig. S 5. The same set of data as shown in Fig. S4 but smeared using gaussian blur to mimic expected experimental resolution.

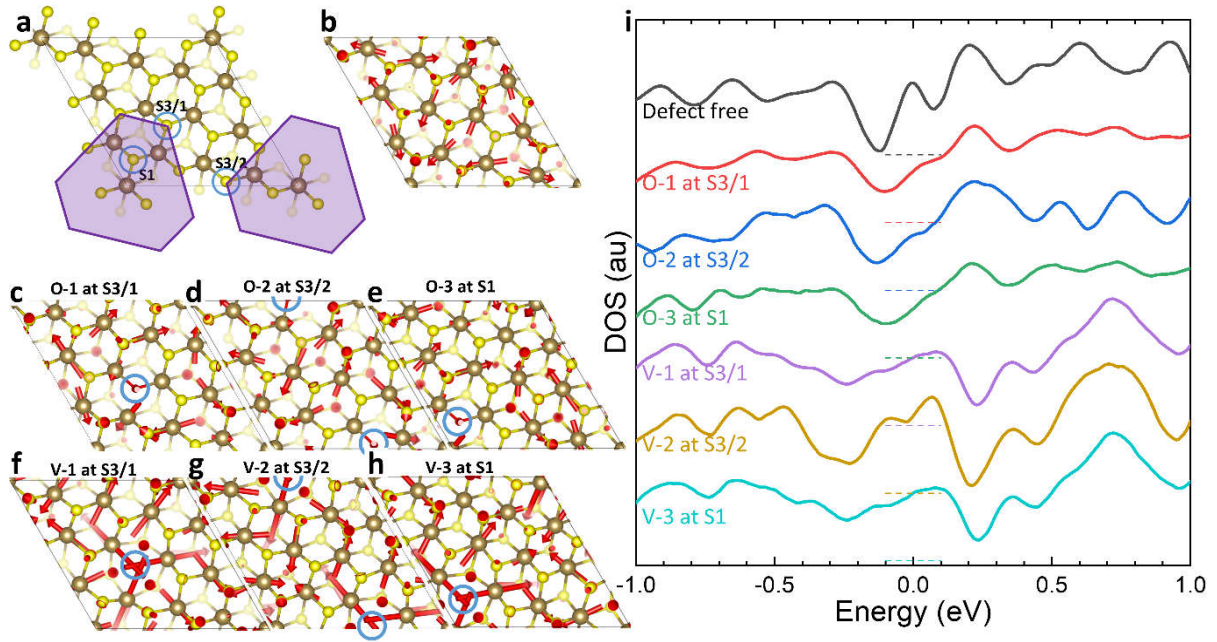

Fig. S 6. (a) The structural DFT model of 1T-TaS$_2$ with indicated three sulfur sites (S1, S3/1, S3/2; for notation see Fig.1 in the main paper) which were used to introduce defects. (b) Structural model with atom deformations as a result of relaxation. (c), (d), (e) show result of structure relaxation after sulfur substitution by oxygen in S3/1, S3/2 and S1 sites respectively. (f), (g), (h) Structural model after introduction of vacancy in in S3/1, S3/2 and S1 sites. Red arrows indicate direction of atoms shifts in respect to not relaxed, PLD-free layer. Lengths of arrows are increased by approx. 10 times to highlight the shifts. (i) DFT calculated DOS showing that qualitatively both oxygen substitution and vacancies show the same trend for each defect type independently on the site at which it (defect) is introduced.

5